# Comment on "Limited surface mobility inhibits stable glass formation for 2-ethyl-1-hexanol" [J. Chem. Phys. 146, 203317 (2017)]

K. L. Ngai[a)] and S. Capaccioli[b)]

[a]CNR-IPCF, Largo Bruno Pontecorvo 3, I-56127, Pisa, Italy

[b]Dipartimento di Fisica, Università di Pisa, Largo Bruno Pontecorvo 3, I-56127, Pisa, Italy

A recent paper by Tylinski et al. [1] reported vapor deposit glasses of 2-ethyl-1-hexanol (2E1H) over a wide range of deposition rates and showed that this molecule does not form highly stable glasses under normal deposition conditions. This abnormality was attributed to the limited surface mobility of 2E1H, which inhibits the formation of its highly stable glasses. Similar deposition rate experiments performed with ethylcyclohexane (ECH) readily forms glasses of high kinetic stability. From this contrast, Tylinski et al. estimated that the surface mobility of 2E1H is more than 4 orders of magnitude less than that of ECH at 0.85 $T_g$. The surface relaxation time, $\tau_{\text{surface}}$, characterizing the surface mobility of 2E1H and ECH were obtained by inferences based upon the stability of vapor-deposited glasses formed, and not from direct measurements such as the grating decay method.

2E1H is a member of the monohydroxy (monoalcohol) molecular glassformers, studied for the first time in the formation of ultrastable glasses. Thus the results on 2E1H of Tylinski et al. are another important contribution to the current research of properties of ultrastable glasses [2-4] as well as enhancement of surface mobility.[5-9]

It is generally agreed that the enhanced mobility at the surface compared to bulk is the key to formation of ultrastable glasses. Three theories have made predictions on the surface mobility [10,11,12]. Tylinski et al. compared with experiment the prediction of the surface relaxation time, $\tau_{\text{surface}}=(\tau_0 \tau_\alpha)^{0.5}$ and a $\tau_0$ value of 1 ps. from the Random First Order Transition (RFOT) theory [10].

Also compared is the prediction of $\tau_{\text{surface}} = (t_c)^n (\tau_\alpha)^{1-n}$ from the CM [12] where $t_c = 2$ ps, $(1-n) \equiv \beta_{\text{KWW}}$ and $\beta_{\text{KWW}}$ is the stretching exponent of the Kohlrausch-Williams-Watts function fitting the α-relaxation. They took for $\beta_{\text{KWW}}$ the value of 0.51 from fits to dielectric loss of 2E1H after subtracting off the more intense Debye relaxation of the monoalcohol at lower frequencies [13]. The Debye relaxation does not contribute to shear modulus[14], and the fit to the fully resolved α-relaxation supports the value of $\beta_{\text{KWW}}$ in the neighborhood of 0.51. The values of $\tau_{\text{surface}}$ calculated by Tylinski et al. from the RFOT and CM are technically correct. However, these values predicted by the two theories are valid only for van der Waals small molecular glass-formers like indomethacin, ortho-terphenyl, and trinapthal benzene. Actually the CM has been applied only to indomethacin[12], ortho-terphenyl and metallic glass [15]. It is inapplicable to polymers at the surface because of it large size,[16] and inapplicable to glass-formers with hydrogen-bonding. Direct measurements of surface diffusion by Yu and coworkers [9] in hydrogen bonded sorbitol, maltitol, and maltose indicate that hydrogen bonding slows down surface diffusion because molecules can optimize hydrogen bonding even near the surface. 2E1H not only has hydrogen-bonding but also the Debye relaxation, which is ascribed by some to the dynamics of clusters formed by hydrogen-bonded molecules [17]. The possibility that the hydrogen bonding can reduce the surface mobility is recognized by Tylinski et al. from the statement they made "Work by Chen and co-workers[48] suggests that the ability to form intermolecular hydrogen bonds may limit the surface mobility and this would be consistent with the slower surface relaxation of 2-ethyl-1-hexanol." Therefore it serves no purpose for anyone to compare the prediction of $\tau_{\text{surface}}$ from either the RFOT or the Coupling Model (CM) with the value inferred from the stability of vapor-deposited glasses of 2E1H such as in Fig.7 of Ref.[1]. Some readers of Ref.[1] may take the comparison in Fig.7, as well as the statement "both predict a very different temperature dependence than what is obtained experimentally", as evidence of failure of the two theories. This is first point of the Comment.

Tylinski et al. also tested the CM prediction of $\tau_{\text{surface}}=(t_c)^n(\tau_\alpha)^{1-n}$ against for ECH by taking $\beta_{\text{KWW}}=0.53$ from dielectric measurements of Mandanici et al.[18,19] The ultrasonic mechanical and dielectric data obtained show a secondary relaxation overlapping the α-relaxation. It has no relation to the α-relaxation because it persists in the liquid state with $\tau_\beta$ <u>longer</u> than the α-relaxation time $\tau_\alpha$, which is unusual for ordinary secondary relaxation, and $\tau_\beta$ continues to have the same Arrhenius T-dependence as in the glassy state. The results suggest that this secondary process is an intramolecular mode overlapping the α-relaxation as recognized by Mandanici et al. The $\tau_\beta$ of ECH is comparable with the relaxation times of the intramolecular relaxation processes resolved by dielectric relaxation and ultrasonics previously found in similar materials containing the cyclohexyl group, including cyclohexane (CH), cyanocyclohexane (CNCH), cyclohexanol (CHOL), and chlorocyclohexane in PS (ClCH:PS) [18,19]. Possibly the intramolecular relaxation observed in ECH and similar glass-formers comes from comformation transitions of the cyclohexyl ring, identified before in poly(cyclohexyl methacrylate.[20,21]

ECH has very low dielectric loss and data were obtained by use of a highly sensitive spectrometer that has a narrow frequency range of 50 Hz–20 kHz less than three decades wide. The overlapping intramolecular mode contributes dielectric loss additively to the α-relaxation with comparable magnitude. This additive contribution makes the fits to the data in the narrow frequency window[18] highly uncertainty in determining the true exponent $\beta_{\text{KWW}}$ of the Kohlrausch-Williams-Watts (KWW) correlation function and coupling parameter n of the α-relaxation. Nevertheless, Mandanici et al. fitted the loss peaks by the Havriliak-Negami equation, and the fit parameters at 112.2 K were converted to yield $\beta_{\text{KWW}}=0.32$. From this and the CM equation, they conclude that the primitive frequency at $T_g$ and hence the JG peak frequency is at $10^{8.4}$ Hz, much faster than any observed dielectric secondary relaxation (see Fig.4 of Ref. [18]). This exercise should not be carried out in the first place because there is no reliable method to account for the broadening of the α-loss peak by the overlapping intramolecular mode at 112.2 K. The results of some related plastic crystals such as cyclohexanol ($\beta_{KWW}$=0.62, or $n$=0.38) by Brand et al.[22] and cyanocyclohexane ($\beta_{KWW}$=0.66, or $n$=0.34) obtained from fit of dielectric loss data of Tschirwitz et al.[23] do not have such a small $\beta_{KWW}$=0.32. In a follow-up report [19] by Mandanici et al. on their study of ECH, horizontal and vertical shifts of isothermal data were applied to compile questionable master curves for the α-relaxation because time-temperature superposition assumed has dubious validity. The revised value of $\beta_{KWW}$=0.53 was obtained from the KWW fit to the master curve. Again, the broadening on both sides of the observed α-loss peak by the conformational transition of the cyclohexyl ring cannot be ignored and the revised $\beta_{KWW}$=0.53 is smaller than the actual value of (1-$n$) where $n$ is the coupling parameter in the CM. Therefore the value of $\beta_{KWW}$=0.53 or $n$=0.47 should not use by Tylinski et al.[1] to calculate $\tau_0=(t_c)^n(\tau_\alpha)^{1-n}$ as the prediction of the CM for ECH and compare with the experimental $\tau_{surface}$ of ECH. This is the second point of the Comment.

Notwithstanding the difficulty in determining the actual coupling parameter $n$ of ECH, there is a way to estimate, or at least to give an upper bound of its value. The chemical structure of ECH differs from cyanocyclohexane (CNCH) in having the ethyl group to replace the compact CN group. The mobile and flexible ethyl group reduces intermolecular coupling in ECH than in CNCH, consistent with the lower $T_g$ equal to 100 K of ECH than 134 K of CNCH.[23] Hence, in accordance with the CM, ECH has smaller value of $n$ than that of CNCH. The faster conformational relaxation in CNCH is well resolved, and its time $\tau_\gamma$ in CNCH is many orders of magnitude shorter than $\tau_\alpha$ at $T_g$ and temperatures above as shown in Fig.1 for $T=141$ K. Consequently, the KWW fit of the isolated α-loss peak at 141 K with $\beta_{KWW}=0.66$ or $n=0.34$ in the figure truly reflects the coupling parameter of CNCH. Indicated by the arrow in Fig.1, the primitive relaxation frequency $\nu_0=(2\pi\tau_0)^{-1}$, obtained from calculating $\tau_0=(t_c)^n(\tau_\alpha)^{1-n}$ with $n=0.34$, is located at the excess wing of the loss data. The latter is the unresolved Johari-Goldstein β-relaxation with relaxation frequency $\nu_{JG}$ of CNCH. Thus the CM prediction of $\nu_0 \approx \nu_{JG}$ is supported by the data of CNCH.

As discussed in the above, we can use the value of $n=0.34$ determined unequivocally for CNCH as the upper bound of the value of $n$ for ECH. We have chosen three values of 0.32, 0.30, and 0.28 and calculate $\tau_0=(t_c)^n(\tau_\alpha)^{1-n}$ from the data of $\tau_\alpha$ and its Vogel-Fulcher fit above $T_g$ as the probable primitive relaxation times of ECH and compare with the experimental $\tau_{surface}$ of ECH in Fig.2. There is approximate agreement between the calculated values of $\tau_0$ and $\tau_{surface}$ obtained by extrapolated the Arrhenius dependence in the glassy state back to $T_g$. The calculated values of $\tau_0$ are weakly dependent on the values of $n$, and this feature helps to support the CM prediction of $\tau_{surface}$ even though the exact value cannot be determined for ECH.

In summary, the RFOT and the CM should not be used for 2E1H by Tylinski et al. to predict $\tau_{surface}$ and compared with the experimental value because the surface mobility is slowed down by hydrogen-bonding and clustering not included in the two theories. For ECH, the CM prediction of $\tau_{surface}=(t_c)^n(\tau_\alpha)^{1-n}$ calculated by using the value of $\beta_{KWW}=0.53$ or $n=0.47$ given by Mandanici et al. is highly questionable due to broadening of the α-relaxation by the overlapping intramolecular mode. The actual value of $n$ is likely smaller had there been no additional contribution from the intramolecular process. Thus the Comment is necessary to rectify any possible misunderstanding of prospective readers of Tylinski et al. concerning these two points.

The ethyl group of ECH makes it more flexible than cyanocyclohexane (CNCH), and thus the coupling parameter $n$ of ECH is smaller than that of CNCH. The value of $n$ equal to 0.34 determined unambiguously for CNCH is then an upper bound of $n$ for ECH. Probable values of $n$=0.32, 0.30, and 0.28 for ECH all give calculated values of $\tau_0=(t_c)^n(\tau_\alpha)^{1-n}$ at and above $T_g$ approximate agreement with $\tau_{\text{surface}}$ inferred from the stability of vapor-deposited glasses. The agreement between $\tau_0$ and $\tau_{\text{surface}}$ is yet another critical test of the CM. This is because $n$ of ECH is significantly smaller than 0.41 for indomethacin [12], 0.50 for OTP and 0.46 for a metallic glass [15], and yet the prediction continues to hold.

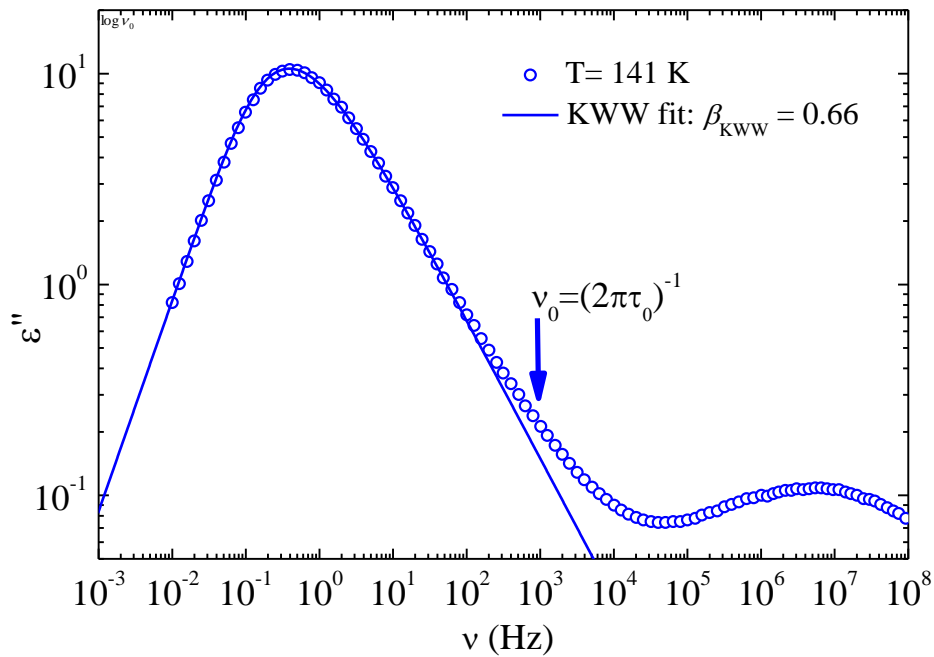

FIG. 1. Dielectric loss spectrum at 141 K of cyanocyclohexane. Data taken from Ref.23. The line is the fit by the Fourier transform of the KWW function with $\beta_{KWW}$=0.66 or $n$=0.34. The resolved fast process is the intramolecular mode. The arrow indicates the calculated primitive relaxation frequency. Its location supports the excess wing is the unresolved JG β-relaxation.

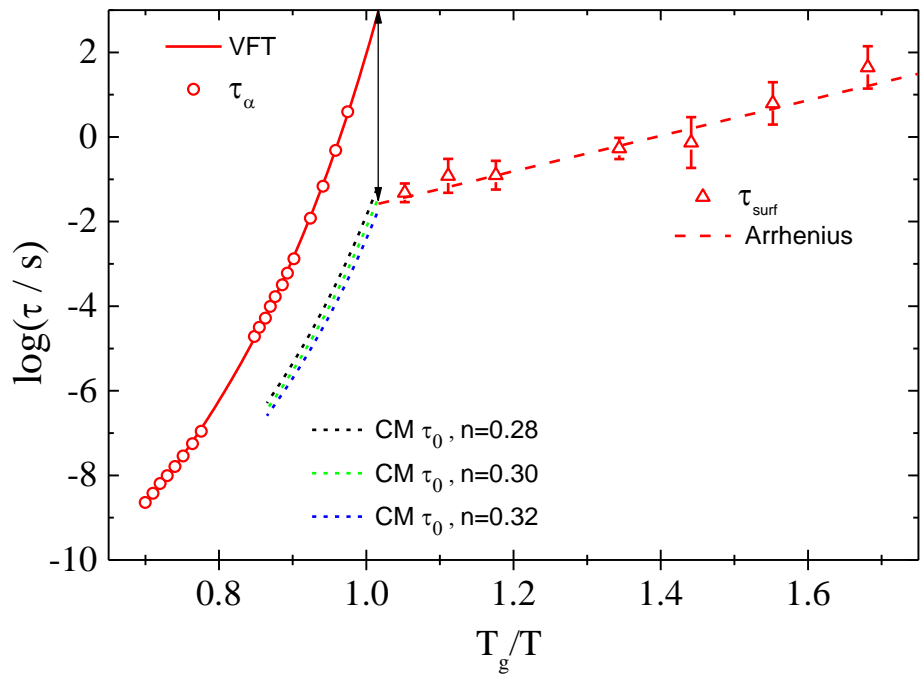

FIG. 2. Relaxation times $\tau_\alpha$ (open circles) and $\tau_{\text{surface}}$ (open triangles) of ethylcyclohexane (ECH) taken from Ref.1. The solid line is the VFT fit of the temperature dependence of $\tau_\alpha$. The dashed line is the fit to $\tau_{\text{surface}}$ by Arrhenius dependence. The three dotted lines represent the calculated primitive relaxation times $\tau_0$ with $n$=0.32, 0.30, and 0.28.